\documentclass[twoside,twocolumn,9pt]{article}
\usepackage{extsizes}
\usepackage[super,sort&compress,comma]{natbib} 
\usepackage[version=3]{mhchem}
\usepackage[left=1.5cm, right=1.5cm, top=1.785cm, bottom=2.0cm]{geometry}
\usepackage{balance}
\usepackage{mathptmx}
\usepackage{sectsty}
\usepackage{graphicx} 
\usepackage{lastpage}
\usepackage{amsmath}
\usepackage{bm}
\usepackage{amssymb}
\usepackage[format=plain,justification=justified,singlelinecheck=false,font={stretch=1.125,small,sf},labelfont=bf,labelsep=space]{caption}
\usepackage{float}
\usepackage{fancyhdr}
\usepackage{fnpos}
\usepackage[english]{babel}
\addto{\captionsenglish}{%
  
}
\usepackage{array}
\usepackage{droidsans}
\usepackage{charter}
\usepackage[T1]{fontenc}
\usepackage[usenames,dvipsnames]{xcolor}
\usepackage{setspace}
\usepackage[compact]{titlesec}
\usepackage{hyperref}

\usepackage{epstopdf}

\definecolor{cream}{RGB}{222,217,201}

\begin{document}

\pagestyle{fancy}
\thispagestyle{plain}
\fancypagestyle{plain}{
\renewcommand{\headrulewidth}{0pt}
}

\makeFNbottom
\makeatletter
\renewcommand\LARGE{\@setfontsize\LARGE{15pt}{17}}
\renewcommand\Large{\@setfontsize\Large{12pt}{14}}
\renewcommand\large{\@setfontsize\large{10pt}{12}}
\renewcommand\footnotesize{\@setfontsize\footnotesize{7pt}{10}}
\makeatother

\renewcommand{\thefootnote}{\fnsymbol{footnote}}
\renewcommand\footnoterule{\vspace*{1pt}%
\color{cream}\hrule width 3.5in height 0.4pt \color{black}\vspace*{5pt}} 
\setcounter{secnumdepth}{5}

\makeatletter 
\renewcommand\@biblabel[1]{#1}            
\renewcommand\@makefntext[1]%
{\noindent\makebox[0pt][r]{\@thefnmark\,}#1}
\makeatother 
\renewcommand{\figurename}{\small{Fig.}~}
\sectionfont{\sffamily\Large}
\subsectionfont{\normalsize}
\subsubsectionfont{\bf}
\setstretch{1.125} 
\setlength{\skip\footins}{0.8cm}
\setlength{\footnotesep}{0.25cm}
\setlength{\jot}{10pt}
\titlespacing*{\section}{0pt}{4pt}{4pt}
\titlespacing*{\subsection}{0pt}{15pt}{1pt}

\fancyfoot{}
\fancyfoot[RO]{\footnotesize{\sffamily{1--\pageref{LastPage} ~\textbar  \hspace{2pt}\thepage}}}
\fancyfoot[LE]{\footnotesize{\sffamily{\thepage~\textbar\hspace{3.45cm} 1--\pageref{LastPage}}}}
\fancyhead{}
\renewcommand{\headrulewidth}{0pt} 
\renewcommand{\footrulewidth}{0pt}
\setlength{\arrayrulewidth}{1pt}
\setlength{\columnsep}{6.5mm}
\setlength\bibsep{1pt}

\makeatletter 
\newlength{\figrulesep} 
\setlength{\figrulesep}{0.5\textfloatsep} 

\newcommand{\topfigrule}{\vspace*{-1pt}%
\noindent{\color{cream}\rule[-\figrulesep]{\columnwidth}{1.5pt}} }

\newcommand{\botfigrule}{\vspace*{-2pt}%
\noindent{\color{cream}\rule[\figrulesep]{\columnwidth}{1.5pt}} }

\newcommand{\dblfigrule}{\vspace*{-1pt}%
\noindent{\color{cream}\rule[-\figrulesep]{\textwidth}{1.5pt}} }

\makeatother

\twocolumn[
  \begin{@twocolumnfalse}
\vspace{1em}
\sffamily
\begin{tabular}{m{4.5cm} p{13.5cm} }

& \noindent\LARGE{\textbf{Singularity identification for the characterization of topology, geometry, and motion of nematic disclination lines}} \\
\vspace{0.3cm} & \vspace{0.3cm} \\

 & \noindent\large{Cody D. Schimming$^{*}$ and Jorge Vi\~nals} \\

\\ & \noindent\normalsize{We introduce a characterization of disclination lines in three dimensional nematic liquid crystals as a tensor quantity related to the so called rotation vector around the line. This quantity is expressed in terms of the nematic tensor order parameter $\mathbf{Q}$, and shown to decompose as a dyad involving the tangent vector to the disclination line and the rotation vector. Further, \textcolor{black}{we derive a kinematic law for the velocity of disclination lines by connecting} this tensor to a topological charge density as in the Halperin-Mazenko description of defects in vector models. Using this framework, analytical predictions for the velocity of interacting line disclinations and of self-annihilating disclination loops are given and confirmed through numerical computation.}

\end{tabular}

 \end{@twocolumnfalse} \vspace{0.6cm}

  ]

\renewcommand*\rmdefault{bch}\normalfont\upshape
\rmfamily
\section*{}
\vspace{-1cm}


\footnotetext{\textit{School of Physics and Astronomy, University of Minnesota, Minneapolis, MN 55455, USA. E-mail: schim111@umn.edu}}





\section{Introduction}
Topological defects play an important role in many physical systems in which a continuous symmetry has been broken. They range from dislocations in crystals, vortices in superconductors, and even cosmic strings in the universe.\cite{chaikin95,kibble97,pismen99} Disclinations in nematic liquid crystals are no exception. Indeed, the observation of disclination lines resulted in the discovery of the nematic phase altogether.\cite{friedel69,deGennes75} In classical (\lq\lq passive") nematics, disclinations are created when domains of mismatching orientation coalesce, or when the boundary conditions---either on the sample itself, or on particles immersed within---disrupt the overall topology of the sample.\cite{deGennes75,kim13,alexander12} An interesting example arises in the \lq\lq Saturn ring" configuration, in which a disclination loop surrounds a particle with homeotropic (i.e. perpendicular) anchoring. \cite{gu00,stark01,alama16}  Another example concerns patterned defects in liquid crystal elastomers which have proven to be a useful means of actuating surfaces.\cite{most15,baba18}

More recently, disclinations in nematics have gained attention in the field of active nematics. In active nematics, the underlying activity causes defects to spontaneously nucleate, and even propels them depending on their geometric character.\cite{marchetti13,ramaswamy17,aranson19} In two-dimensions, topological defects have been observed as points of interest in microtubule systems, bacterial suspensions, soil bacteria, and epithelial tissue.\cite{kumar18,genkin17,nishiguchi17,copenhagen21,saw17} In three-dimensions, recent experimental and computational work on microtubule systems shows the nucleation, active flow, and eventual annihilation or recombination of lines and loops.\cite{duclos20} Theoretical and computational work has aided in understanding how the various geometric features of disclination loops result in differing flow patterns.\cite{binysh20,houston21}

There have also been recent efforts to characterize disclination lines. Long \textit{et al.}\cite{long21} have shown that the geometric properties of disclination lines can be expressed through a series of tensors from ranks 1--3. These properties determine the force of one line on another, as well as their active flow. Additionally, other investigations\cite{shankar19,angheluta21} have characterized disclinations in two-dimensions as particles, and connected their velocity to a conserved topological current density. These characterizations have important implications for identifying defect positions and velocities in both experimental systems and numerical computations. Further, they shift the perspective of predicting liquid crystal behavior to defects, which in some cases allows analytic calculation. 

\begin{figure}[h]
\centering
  \includegraphics[height=6cm]{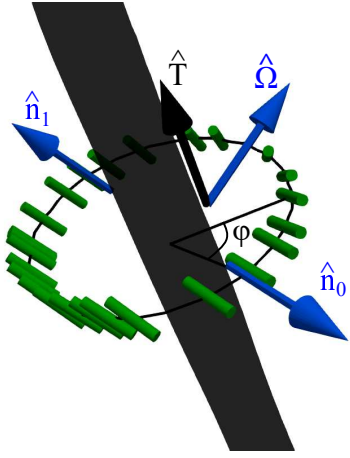}
  \caption{Schematic example of a disclination line showing its geometric features. $\mathbf{\hat{T}}$ is the unit tangent vector and $\{\mathbf{\hat{n}_0},\,\mathbf{\hat{n}_1},\,\bm{\hat{\Omega}}\}$ describes the orientation of the nematogens (depicted as cylinders) as they encircle the disclination core.}
  \label{fgr:DiscExample}
\end{figure}

To discuss the topological character of defects in nematic liquid crystals, one starts with the order parameter symmetry---or, more precisely, the ground state manifold---namely \textcolor{black}{the real projective space, $\mathbb{RP}^{n-1}$}. \textcolor{black}{More generally}, systems which break $SO(n)$ \textcolor{black}{(rotational)} symmetry can be represented by an n-dimensional vector which goes to zero at defect locations, and $d-n$, where $d$ is the spatial dimension, determines the dimension of the topological charge density.\cite{liu92,mazenko97} For example, in two-dimensional nematics the topological charge density is a scalar because the ground state manifold is \textcolor{black}{$\mathbb{RP}^1 \cong SO(2)$} and the order parameter can be represented by a complex number.\cite{angheluta21} The situation in three-dimensions is more complex. Although the order parameter \textcolor{black}{breaks three dimensional rotational symmetry ($SO(3)$)}, the extra cylindrical and apolar symmetries \textcolor{black}{determine the ground state manifold to be $\mathbb{RP}^2$ which} allows both line defects and point defects.\cite{alexander12} Thus, two types of topological charge \textcolor{black}{densities} exist: a scalar for point defects and a vector for line defects \textcolor{black}{(see Eq. \eqref{eqn:DiscDensT})}. Further complexities arise because there is only one topological equivalency class of line defects (as opposed to infinite half-integral charges in two-dimensions). Geometrically this appears as a ``rotation vector,'' $\boldsymbol{\hat{\Omega}}$.\cite{friedel69,duclos20} This rotation vector defines the plane the nematogens lie in as they encircle the disclination. Fig. \ref{fgr:DiscExample} shows a general example of a disclination with rotation vector $\boldsymbol{\hat{\Omega}}$, and unit tangent vector $\mathbf{\hat{T}}$. The cylinders outside the disclination represent the orientation of nematogens.

In this work, we extend these recent efforts to characterize disclination lines by defining a disclination density tensor, valid in three-dimensions. Our primary result is the definition of this tensor as a function of first derivatives of the nematic tensor order parameter, $\mathbf{Q}$, which is typically the preferred representation of the nematic near defects.\cite{mottram14} We further show that this disclination density tensor has a simple decomposition in terms of the rotation vector $\boldsymbol{\hat{\Omega}}$ and the unit tangent vector $\mathbf{\hat{T}}$, and can be used to identify and classify disclination lines. This disclination density tensor is then used as a starting point to discuss the dynamics of disclination lines.

In Section 2, we define the disclination density tensor and examine its relationship to the geometric properties of a disclination. We also present several numerical realizations, demonstrating the utility of this characterization. In Section 3 we \textcolor{black}{derive a kinematic law for the velocity of disclination lines by} connecting \textcolor{black}{the disclination density tensor} to the topological charge density of disclinations, and invoke the Halperin-Mazenko singularity tracking method\cite{halperin81,liu92} to derive continuity equations for the topological charge density in terms of $\mathbf{Q}$. We show that the velocity of disclination lines is dependent on derivatives of $\mathbf{Q}$ at the defect core. These are kinematic results that hold regardless of any assumption on the \textcolor{black}{dynamic law governing the} time evolution of $\mathbf{Q}$, and so they hold for both passive and active nematics with mass transport. In Section 4 we use this framework and a linear approximation of $\mathbf{Q}$ to analytically predict the velocity for interacting disclination lines, and self-annihilating passive loops as a function of their radii. We confirm the predictions with three-dimensional numerical solutions for the evolution of $\mathbf{Q}$, and show that the result for interacting disclination lines is equivalent to that of Long \textit{et al.}\cite{long21} for the Peach-Koehler force between two disclinations.

\section{Disclination density tensor}
\subsection{Definitions}
In two dimensional nematics, topological defects are point-like singularities that can be described by a closed curve $C$ encircling the singularity. In particular, the charge, $m$, of the defect is defined by
\begin{equation} \label{eqn:2Ddefect}
    2 \pi m = \oint_{C} \varepsilon_{\mu \nu} \hat{n}_{\mu} \partial_k \hat{n}_{\nu} \, d \ell_k
\end{equation}
where $\mathbf{\hat{n}}$ is the representative vector of local order, called the director. We denote by $\partial_k$ the derivative $ \partial / \partial x_k$, and summation over repeated indices is assumed. It is simple to verify that the integrand $\varepsilon_{\mu \nu} \hat{n}_{\mu} \partial_k \hat{n}_{\nu} = \partial_k \theta$ where $\theta$ is the angle the director makes with some reference axis. Because of the apolar symmetry in nematic liquid crystals, half-integer defects are allowed and $m = \pm 1/2$ are the lowest energy configurations containing a defect.\cite{deGennes75} Upon combining, these defects add their charges as rational numbers.

In three-dimensions, topological defects in nematics manifest as both lines and points. The first objective of this paper is to classify the lines in a way that can be computed directly from the order parameter. This has been previously done for point defects, up to a sign ambiguity.\cite{alexander12} However, to our knowledge, there is not a generalization of Eq. \eqref{eqn:2Ddefect} for line disclinations. This is because the topology of lines in three-dimensional nematics is different. All line disclinations in nematics have a charge of $+1/2$, but upon combining they add as elements of the group $\mathbb{Z}_2$. That is, any two line defects that combine will annihilate each other, even if they energetically repel.

The first step to generalizing Eq. \eqref{eqn:2Ddefect} for line disclinations is to map its director field to an equivalent two-dimensional vector field. This is required because in order to have line defects in a material with rotational symmetry breaking, the dimension of the order parameter must be one dimension lower than that of the ambient space.\cite{mazenko97} In Fig. \ref{fgr:DiscExample}, the director field around the disclination in its normal plane is given by
\begin{equation} \label{eqn:nDisc}
    \mathbf{\hat{n}} = \mathbf{\hat{n}_0}\cos \frac{1}{2} \varphi + \mathbf{\hat{n}_1} \sin \frac{1}{2} \varphi
\end{equation}
where $\{\mathbf{\hat{n}_0},\mathbf{\hat{n}_1},\boldsymbol{\hat{\Omega}}\}$ is an orthonormal triad of vectors, and $\varphi$ represents the azimuthal angle around the disclination in its normal plane with respect to some reference axis. This parametrization serves as an intuitive geometric definition for $\boldsymbol{\hat{\Omega}}$: $\bm{\hat{\Omega}} \cdot \mathbf{\hat{n}} = 0$ as $\mathbf{\hat{n}}$ circles the disclination. However, finding this vector at a point on the disclination from a measuring circuit $C$ is more difficult. 

\begin{figure}[h]
\centering
  \includegraphics[height=4in]{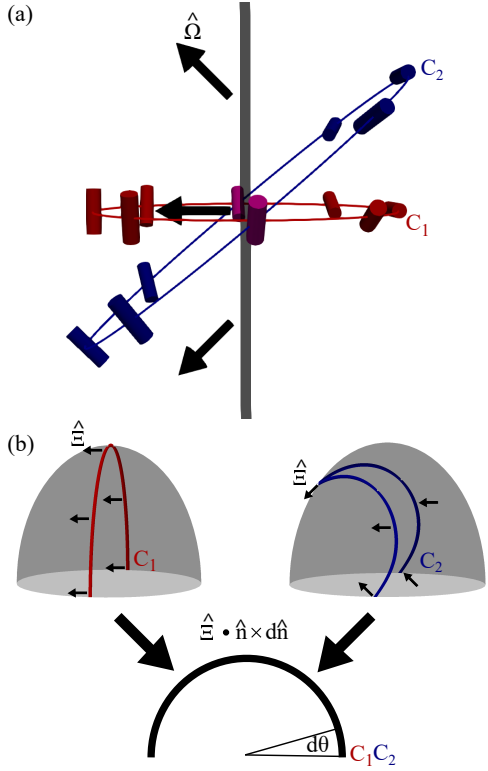}
  \caption{The charge of a disclination with $\bm{\hat{\Omega}}$ varying along the line is measured with two curves, $C_1$ and $C_2$. (a) $C_1$ remains in the normal plane of the disclination where $\bm{\hat{\Omega}}$ is \textcolor{black}{well-defined}, while $C_2$ is out of the normal plane. (b) Curves $C_1$ and $C_2$ in order parameter space. \textcolor{black}{$\bm{\hat{\Xi}}$} is defined as the vector orthogonal to both $\mathbf{\hat{n}}$ and its derivative along the curve. Direct integration over the paths in order parameter space yield different results. However integrating the projected $\textcolor{black}{\bm{\hat{\Xi}}} \cdot \mathbf{\hat{n}} \times d\mathbf{\hat{n}}$ yields the correct charge for both curves since the projection collapses both curves onto the half circle with ends identified.}
  \label{fgr:OmegaProject}
\end{figure}

To exemplify this difficulty we show in Fig. \ref{fgr:OmegaProject} a disclination line with $\bm{\hat{\Omega}}$ changing along the line. Two curves, $C_1$ and $C_2$ are used to measure the defect charge. $C_1$ remains in the normal plane of the disclination and as such the corresponding curve in order parameter space, shown in Fig. \ref{fgr:OmegaProject}b, is planar. On the other hand, $C_2$ is out of the normal plane and so its order parameter equivalent is also out of the plane. Naively integrating $d\mathbf{\hat{n}}$ along this curve will not yield the same result, i.e. the integral is path dependent. Of course, topologically, these two curves differ only by a continuous transformation and are equivalent; however, $\bm{\hat{\Omega}}$ is not \textcolor{black}{well-defined} along the second curve. The resolution to this paradox is that the charge of a disclination line is a scalar, and $\bm{\hat{\Omega}}$ is not a topological invariant of the disclination line. \textcolor{black}{Instead, we can measure the charge by defining a local vector away from the defect core.} Since $\mathbf{\hat{n}}$ is a unit vector, its derivative in some direction will be orthogonal to itself. Therefore, for a measuring circuit $C$, we can define the vector \textcolor{black}{$\bm{\hat{\Xi}}$} locally as the vector that is orthogonal to both $\mathbf{\hat{n}}$ and its derivative along the curve, \textcolor{black}{which we will denote $d \mathbf{\hat{n}}$. Then $\bm{\hat{\Xi}} \equiv \mathbf{\hat{n}} \times d \mathbf{\hat{n}} / |\mathbf{\hat{n}} \times d \mathbf{\hat{n}}|$ at each point along a chosen measuring curve.} To integrate along the curves in order parameter space, we must \textit{locally} project into this orthogonal direction. This has the effect of mapping a curve in order parameter space to the half circle with end points identified, as seen in Fig. \ref{fgr:OmegaProject}b. Thus, after taking the result modulo $2\pi$ we can \textcolor{black}{compute} the disclination charge through the following relation:
\begin{equation} \label{eqn:OmegaDef}
    \textcolor{black}{\pi p}  = \oint_C \textcolor{black}{\hat{\Xi}_{\gamma}} \varepsilon_{\gamma \mu \nu} \hat{n}_{\mu} \partial_k \hat{n}_{\nu} \, d \ell_k
\end{equation}
where $\textcolor{black}{p} \in \{0,\,1\}$ \textcolor{black}{is computed modulo $2$ indicating the charge $m = p / 2$}, and \textcolor{black}{$\bm{\hat{\Xi}}$} is not necessarily a constant vector but is defined by $\mathbf{\hat{n}}$ and its derivative locally. \textcolor{black}{The integrand of Eq. \eqref{eqn:OmegaDef} measures the rate of rotation of $\mathbf{\hat{n}}$ along the curve, though we keep this explicit form of projecting into the vector $\bm{\hat{\Xi}}$ since it is useful in both conceptualization and mathematical brevity in what follows.} \textcolor{black}{As the measuring circuit is taken to be smaller and smaller, $\bm{\hat{\Xi}} \to \bm{\hat{\Omega}}$, thus} it is useful to identify $\bm{\hat{\Omega}}$ \textcolor{black}{as a property of the} defect core since the director nearby can be approximated by Eq. \eqref{eqn:nDisc}. However, this approximation may break down far from the defect due to curvature of the defect, other defects, and boundary effects. Therefore, for arbitrary measuring circuits, it is important to define \textcolor{black}{$\bm{\hat{\Xi}}$} along the curve chosen as in Eq. \eqref{eqn:OmegaDef}.

Equation \eqref{eqn:OmegaDef} is the three-dimensional generalization of Eq. \eqref{eqn:2Ddefect}, however, it assumes knowledge of \textcolor{black}{$\bm{\hat{\Xi}}$} everywhere and so is not practically very useful. We use this relation as a starting point to derive our first primary result. First, though, we briefly note that the integrands in Eqs. \eqref{eqn:2Ddefect} and \eqref{eqn:OmegaDef} are similar to the effective strain used by Long \textit{et al.} to define the effective Peach-Koehler force on disclination lines \cite{long21}. This is no accident: taking $\mathbf{\hat{n}}$ as in Eq. \eqref{eqn:nDisc} we find $\hat{\Omega}_{\gamma} \varepsilon_{\gamma \mu \nu} \hat{n}_{\mu} \partial_k \hat{n}_{\nu} = (1 / 2) \partial_k \varphi$. For a two-dimensional nematic, the effective disclination strain is $m \nabla \varphi$ where $m$ is the disclination charge. Hence the similarity in the expressions. 

In the presence of disclinations, it is typically advantageous to represent the nematic with a tensor order parameter, $\mathbf{Q}$, as opposed to a singular vector. $\mathbf{Q}$ regularizes the singularity at the center of the defect, and remains continuous for half integer defects (note Eq. \eqref{eqn:nDisc} reverses sign for $\varphi = 2 \pi$). We parametrize $\mathbf{Q}$ by $\mathbf{Q} = S \left[ \mathbf{\hat{n}} \otimes \mathbf{\hat{n}} - (1 / 3) \mathbf{I} \right] + P \left[ \mathbf{\hat{m}} \otimes \mathbf{\hat{m}} - \bm{\hat{\ell}} \otimes \bm{\hat{\ell}} \right]$ where $\{\mathbf{\hat{n}},\mathbf{\hat{m}},\boldsymbol{\hat{\ell}}\}$ are an orthonormal triad, and $\mathbf{\hat{n}}$ is the director. $S$ and $P$ parametrize the eigenvalues of $\mathbf{Q}$, and represent the degree of ordering of the nematogens. Specifically, $S$ represents uniaxial order and $P$ represents biaxial order. Although we focus here on uniaxial liquid crystals, it is known that the distribution of nematogens near the core of disclinations becomes biaxial.\cite{schopohl87} However, at the center of a disclination, two of the eigenvalues of $\mathbf{Q}$ cross and $S-P = 0$.

\begin{figure*}[htbp]
\centering
  \includegraphics[height=8in]{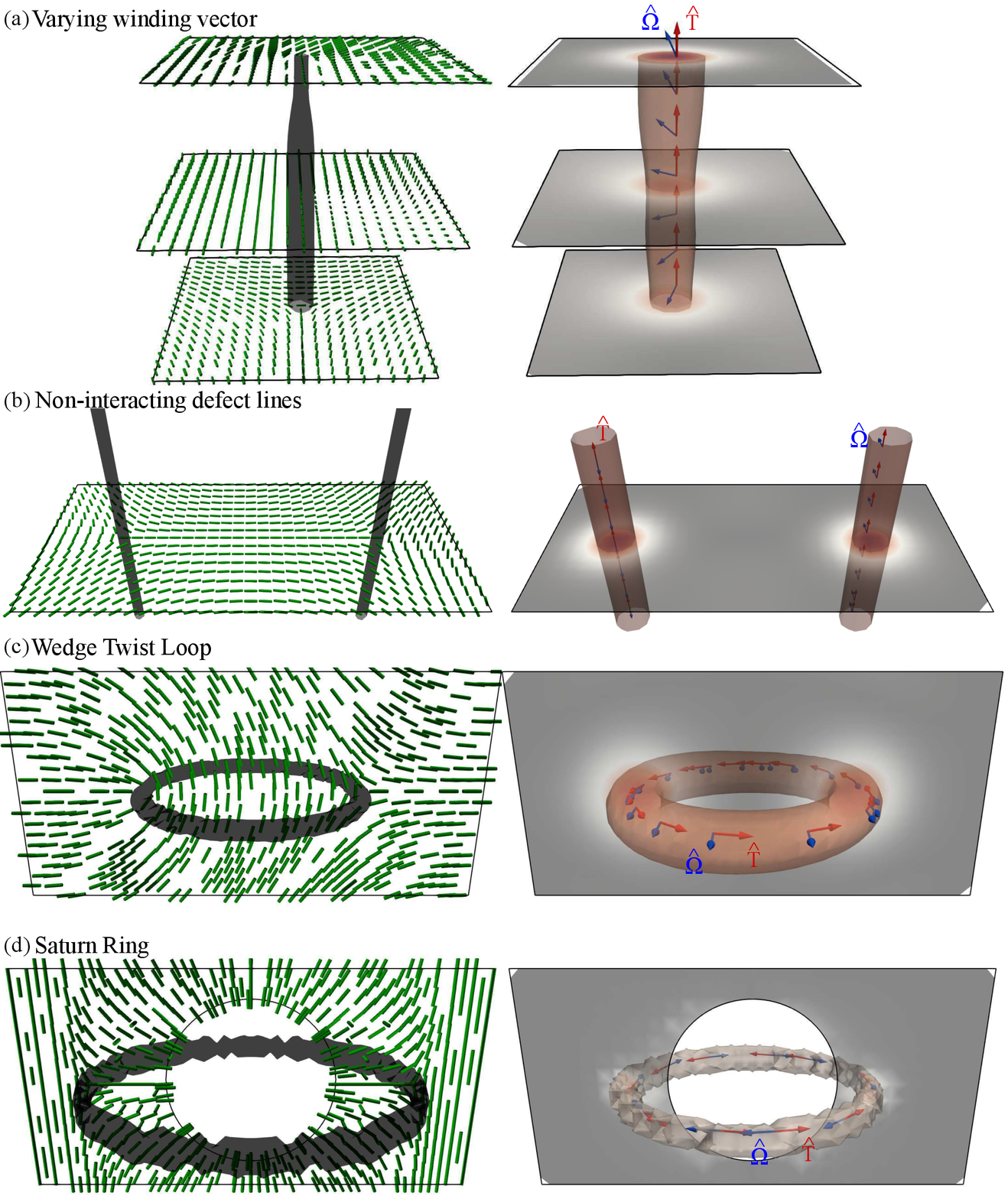}
  \caption{Various defect configurations (left) with the computed decomposition of the disclination density tensor $\mathbf{D}$, Eq. \eqref{eqn:DDecomp} (right). Nematic configurations are computed using a finite element gradient flow algorithm with a Maier-Saupe bulk free energy and a one-constant elastic free energy. In all figures, green cylinders represent nematogen orientations; black contours represent the defect core where the scalar order parameter $S = 0.3 S_{0}$; black to red color map shows $\omega(\mathbf{r})$ with red contours showing where $\omega = 0.7\omega_{\text{max}}$; blue arrows show $\bm{\hat{\Omega}}$; and red arrows show $\mathbf{\hat{T}}$. (a) A line defect with varying rotation vector $\bm{\hat{\Omega}}$. (b) Non-interacting disclination lines with orthogonal rotation vectors. (c) A wedge-twist disclination loop with constant rotation vector but varying tangent vector. (d) A Saturn-ring configuration with rotation vector anti-parallel to the tangent vector at all points.}
  \label{fgr:TopExamples}
\end{figure*}

We now extend Eq. \eqref{eqn:OmegaDef} in terms of the tensor order parameter. To accomplish this we first assume we are working far enough from the defect so that the distribution of nematogens is uniaxial, that is $P=0$. We restrict our measuring curve, $C$, to only pass through points of constant $S = S_0$ and we will denote this curve as $C_0$. Then, derivatives of $\mathbf{Q}$ are equivalent to derivatives of $\mathbf{\hat{n}}$ and it can be shown that the charge integral in terms of $\mathbf{Q}$ is 
\begin{equation} \label{eqn:OmegaQ}
    \textcolor{black}{S_0^2 \pi p}   = \oint_{C_0} \textcolor{black}{\hat{\Xi}_{\gamma}} \varepsilon_{\gamma \mu \nu} Q_{\mu \alpha} \partial_k Q_{\nu \alpha} \, d \ell_k.
\end{equation}
This generalizes Eq. \eqref{eqn:OmegaDef}. However, because we must work away from defects where $S$ is constant it is not practically useful. To construct a more useful quantity, we apply Stoke's theorem to Eq. \eqref{eqn:OmegaQ}. This yields the first main result of this paper:
\begin{equation} \label{eqn:DDef}
    \textcolor{black}{S_0^2 \pi p} = \int_{\Gamma_0} \textcolor{black}{\hat{\Xi}_{\gamma}} \varepsilon_{\gamma \mu \nu} \varepsilon_{i \ell k} \partial_{\ell} Q_{\mu \alpha} \partial_k Q_{\nu \alpha} \, d a_i \equiv \int_{\Gamma_0} \textcolor{black}{\hat{\Xi}_{\gamma}} D_{\gamma i} \, d a_i
\end{equation}
where $\Gamma_0$ is a surface bounded by curve $C_0$ and $d a_i$ is an element of area on the surface. Note that in taking the curl of the integrand of Eq. \eqref{eqn:OmegaQ} there should be three terms since we are assuming \textcolor{black}{$\bm{\hat{\Xi}}$} is spatially varying as well. However, since \textcolor{black}{$\bm{\hat{\Xi}}$} is a unit vector, its derivative will be orthogonal to itself and thus we can move \textcolor{black}{$\bm{\hat{\Xi}}$} out of the derivative. The third term is also zero since it is the curl of the gradient of $\mathbf{Q}$ which is not a singular quantity. Equation \eqref{eqn:DDef} serves as a definition of the tensor $\mathbf{D}$ which we call the ``disclination density tensor.'' As a check, for two-dimensional systems the appropriate quantity is $D_{3 3} = \varepsilon_{3\mu\nu}\varepsilon_{3 \ell k} \partial_{\ell} Q_{\mu \alpha} \partial_k Q_{\nu \alpha}$. This quantity has been used in other investigations to track and identify defects in two-dimensional (and quasi two-dimensional) systems.\cite{blow14,dell18} Therefore the generalized tensor, $\mathbf{D}$, goes to the appropriate scalar in the two dimensional limit.

\subsection{Properties of D}
We now delineate some useful properties of $\mathbf{D}$. First, $\mathbf{D}$ is a smooth tensor field, owing to the regularization that $\mathbf{Q}$ provides. Further, $\mathbf{D} = \mathbf{0}$ where $S$ is constant which can be seen by substituting our parametrization of $\mathbf{Q}$ into the definition of $\mathbf{D}$, Eq. \eqref{eqn:DDef}. Therefore the points where $\mathbf{D} \neq \mathbf{0}$ mark disclinations.

For disclination lines, $\mathbf{D}$ decomposes nicely as
\begin{equation} \label{eqn:DDecomp}
    \mathbf{D}(\mathbf{r}) = \omega (\mathbf{r}) \left( \bm{\hat{\Omega}} \otimes \mathbf{\hat{T}} \right)
\end{equation}
where $\omega(\mathbf{r})$ is a non-negative scalar field which is at its maximum at the disclination core, and $\mathbf{\hat{T}}$ is the disclination line tangent vector. This decomposition can be seen immediately if we take $\omega(\mathbf{r}) = \delta(\mathbf{r} - \mathbf{R})$ where $\mathbf{R}$ is the location of the defect line and substitute Eq. \eqref{eqn:DDecomp} into Eq. \eqref{eqn:DDef}. The delta-function expression for $\omega$ is valid for a singular field such as $\mathbf{\hat{n}}$, though $\omega$ smooths out to the size of the core for a regularized field like $\mathbf{Q}$. \textcolor{black}{We also find that $\omega$ goes to zero at the core of integer line defects where, in three dimensions, the escape to the third dimension\cite{deGennes75} destroys the linear character of the defect.}

\textcolor{black}{Another useful property of $\mathbf{D}$ is that it inherently fixes the sign of $\bm{\hat{\Omega}}\cdot\mathbf{\hat{T}}$. A common issue with determining the character of a disclination line is that the independent vectors $\bm{\hat{\Omega}}$ and $\mathbf{\hat{T}}$ are defined only up to a sign and it is the sign of their scalar product that determines the winding character of the disclination. For $\mathbf{D}$, this scalar product is proportional to its trace and, hence, once a direction for $\mathbf{\hat{T}}$ (or $\bm{\hat{\Omega}}$) is chosen the sign of the other vector is fixed by definition. Thus, if one is only interested in the winding character of a disclination line, one needs only to compute the trace of $\mathbf{D}$.}

To demonstrate the practical usefulness of the decomposition, Eq. \eqref{eqn:DDecomp}, we show in Fig. \ref{fgr:TopExamples} several examples of disclination lines and loops, alongside their respective $\omega, \, \bm{\hat{\Omega}}$, and $\mathbf{\hat{T}}$ fields. To compute these, one must have access to the first derivatives (or numerical equivalents) of the order parameter, $\mathbf{Q}$. Then, using Eq. \eqref{eqn:DDef}, one can compute $\mathbf{D}$. $\omega$ is computed as the Frobenius norm of $\mathbf{D}$ while $\bm{\hat{\Omega}}$ ($\mathbf{\hat{T}}$) is the non-degenerate eigenvector of $\mathbf{D} \mathbf{D}^T$ ($\mathbf{D}^T \mathbf{D}$). \textcolor{black}{The final step is to ensure that both $\mathbf{\hat{T}}$ and $\bm{\hat{\Omega}}$ are continuous along the disclination line, which can be done by fixing the direction of the tangent line and then fixing $\bm{\hat{\Omega}}$ by $\text{sgn}\left(\bm{\hat{\Omega}}\cdot\mathbf{\hat{T}}\right) = \text{sgn}(\text{Tr}\mathbf{D})$.} The examples were numerically computed using a Maier-Saupe bulk free energy with a ``one-constant'' Landau-de Gennes elastic free energy (more details on the computations are given below in Section 4). 

In Fig. \ref{fgr:TopExamples}a we show the ``counter-example'' for the normal plane measuring circuit also shown in Fig. \ref{fgr:OmegaProject}. As evidenced by the figure, the decomposition of $\mathbf{D}$ picks out the value of $\bm{\hat{\Omega}}$ that changes along the line. Fig. \ref{fgr:TopExamples}b shows two disclination lines with orthogonal rotation vectors, which were shown by Long \textit{et al.}\cite{long21} to be non-interacting. Fig. \ref{fgr:TopExamples}c shows a snapshot of a self-annihilating wedge-twist loop disclination,\cite{duclos20} showing that the decomposition is just as useful for curved disclinations. Finally, Fig. \ref{fgr:TopExamples}d shows a Saturn ring\cite{gu00,alama16} configuration where homeotropic anchoring on a colloidal particle topologically requires the existence of a disclination loop with rotation vector anti-parallel to the tangent vector. 

\begin{figure}[h]
\centering
  \includegraphics[width=3.3in]{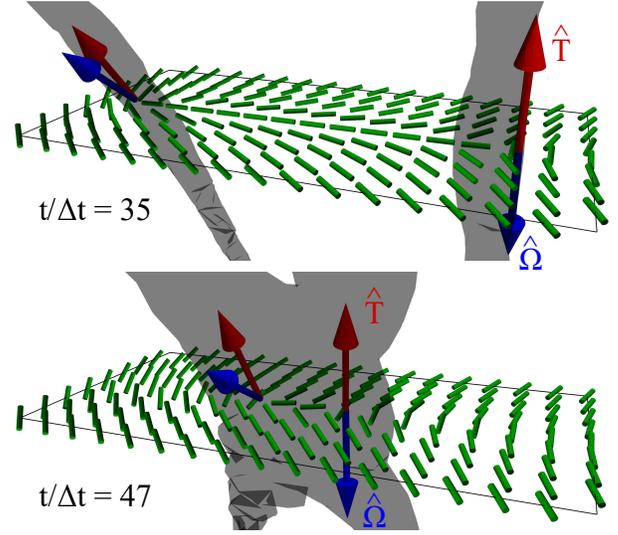}
  \caption{\textcolor{black}{Defect geometry near the recombination of disclination lines with skewed initial tangent and rotation vectors. The configurations at two separate iteration numbers are shown, $t/\Delta t = 35,\, 47$. The green cylinders represent the nematogen orientation, the black contours represent the defect core where the scalar order parameter $S_0 = 0.2 S$, the blue arrows show $\bm{\hat{\Omega}}$, and the red arrows show $\mathbf{\hat{T}}$. As the lines recombine at the closest point between them, the tangent vectors rotate to be nearly parallel, however the rotation vectors show little change.}}
  \label{fgr:DefectRecoClose}
\end{figure}

Because we only need to compute the first derivative of the tensor order parameter, this method of identifying defects and obtaining geometric information is powerful, and should prove useful, particularly in studies of active nematic systems in three-dimensions in which defect lines and loops are spontaneously nucleated and annihilated.\cite{duclos20,binysh20} \textcolor{black}{To exemplify this we show in Fig. \ref{fgr:DefectRecoClose} the disclination line geometry computed from $\mathbf{D}$ as two disclination lines with skewed tangent vectors and rotation vectors are close to annihilating. This example is more complex than those shown in Fig. \ref{fgr:TopExamples} since the annihilation of the disclination lines causes changes in the curvature of the defects and rotation of the tangent vectors (see Fig. \ref{fgr:LineRecombo} as well). The tangent vector rotation can be seen from comparing the two plots in Fig. \ref{fgr:DefectRecoClose}. Additionally, we find that the rotation vectors, $\bm{\hat{\Omega}}$, rotate little in the annihilation process.}

\textcolor{black}{While we have shown both simple and complex examples of computation of the disclination density tensor for weakly curved defects, we do not expect this construction to hold for strongly curved defects such as in the transient stages of defect nucleation. By strongly curved, we mean that $\kappa a \sim 1$ where $\kappa$ is the defect curvature and $a$ is the radius of the defect core. For curvatures this large, the continuum description of the disclination breaks down and the construction of $\mathbf{D}$ is no longer valid.}

\textcolor{black}{To conclude this section we comment on the methods for determining $\bm{\hat{\Omega}}$ laid out in the supplementary information of Ref.\cite{duclos20} and how they compare to our methods presented above. First, the local formula, $\bm{\tilde{\Omega}} = \mathbf{\hat{n}} \times \left(\mathbf{\hat{n}} \cdot \nabla \right) \mathbf{\hat{n}}$ is similar to the definition of $\bm{\hat{\Xi}}$, except that there is no reference to a measuring curve and the directional derivative is in the direction of $\mathbf{\hat{n}}$ rather than in the direction of the curve. Thus $\bm{\tilde{\Omega}}$ is proportional to $\bm{\hat{\Omega}}$ at the disclination core since $\bm{\hat{\Xi}}\to\bm{\hat{\Omega}}$ but goes to zero for pure twist disclinations. The other method is a non-local construction of the curve in order parameter space where $\mathbf{\hat{n}}$ near the disclination core is extracted along a curve in the normal plane to the disclination and $\bm{\hat{\Omega}}$ is the normal vector to the curve in order parameter space. This gives the correct $\bm{\hat{\Omega}}$ as long as $\mathbf{\hat{n}}$ is as in Eq. \eqref{eqn:nDisc} (see the curve $C_1$ in Fig. \ref{fgr:OmegaProject}). However, a local formula is more desirable since $\mathbf{\hat{n}}$ may deviate from Eq. \eqref{eqn:nDisc} due to external constraints. Hence, the disclination density tensor represents a local construction that can robustly determine the character of a disclination line and should prove useful for future investigations.} 

\section{\textcolor{black}{Kinematic} equations for disclination lines}
We now \textcolor{black}{derive a kinematic equation for the velocity of disclination lines which holds regardless of details of the dynamical model chosen for $\mathbf{Q}$.} \textcolor{black}{We first} connect the tensor $\mathbf{D}$ to the Jacobian appearing in the Halperin-Mazenko formalism of topological defects in systems with $SO(n)$ symmetry.\cite{liu92,mazenko97,mazenko99} The appropriate density for disclination lines is\cite{liu92}
\begin{equation} \label{eqn:DiscDensT}
    \bm{\rho}(\mathbf{r}) = \frac{1}{2} \sum_j\int \frac{d \mathbf{R}_j}{d s} \delta[\mathbf{r} - \mathbf{R}_j(s)] \, d s
\end{equation}
where $\mathbf{R}_j(s)$ is the $j$th disclination line's position at point $s$ along the curve. In the Halperin-Mazenko formalism, one connects this line defect density to order parameter space through a ``zero-finding'' delta function of a complex order parameter (i.e. one whose amplitude goes to zero at defect locations).

In two dimensions, one searches for zeros of $\mathbf{Q}$ and the defect density is related to $\delta[\mathbf{Q}]$ via the Jacobian\cite{angheluta21} $(1/2)\varepsilon_{\mu \nu} \varepsilon_{\ell k} \partial_{\ell} Q_{\mu \alpha} \partial_k Q_{\nu \alpha}$. In three dimensions, however, $\mathbf{Q} \neq 0$ at the location of defects. Instead, two of its three eigenvalues cross, and $S - P = 0$ at the core. Additionally, since $\mathbf{n}$ is orthogonal to $\bm{\hat{\Omega}}$ near the core, there is only one degree of freedom describing its orientation. Hence, instead of looking for zeros of $\mathbf{Q}$, we search for zeros in a two dimensional subspace of the full five dimensional order parameter space. This subspace is defined by $S-P$ and the orientation of the nematogen. We denote the corresponding delta function symbolically as $\delta[\mathbf{Q}_{\perp}(\mathbf{r})]$.

Now it remains to calculate the Jacobian. Since the space we are working with is two dimensional, we may work in polar coordinates in order parameter space. We take our radial component to be $(S-P)^2$ while the azimuthal component is $\theta$, representing the orientation of the nematogens. Note that near the core, the integrand of Eq. \eqref{eqn:OmegaQ} can be written in terms of $(S-P)^2$ and $\theta$:
\begin{equation} \label{eqn:polarQ}
    \hat{\Omega}_{\gamma} \varepsilon_{\gamma \mu \nu} Q_{\mu \alpha} \partial_k Q_{\nu \alpha} = (S - P)^2 \partial_k \theta.
\end{equation}
In the typical polar coordinate representation, the Jacobian can be computed as $(1/2)\nabla \times (r \nabla \varphi)$. Hence, identifying $(S-P)^2 \equiv r$, the Jacobian of our subspace is the curl of Eq. \eqref{eqn:polarQ}. Comparing with Eq. \eqref{eqn:DDef}, \textcolor{black}{and noting that we are working near the core so we can substitute $\bm{\hat{\Xi}}$ with $\bm{\hat{\Omega}}$, the Jacobian} is $(1/2) \bm{\hat{\Omega}}\cdot \mathbf{D}$. We note that this expression reduces to the correct Jacobian in two dimensions.

With this expression for the Jacobian, the disclination density can be written in terms of the order parameter,\cite{mazenko97}
\begin{equation} \label{eqn:DiscDensD}
    \rho_i(\mathbf{r}) = \frac{1}{2} \delta \left[ \mathbf{Q}_{\perp}(\mathbf{r})\right] \hat{\Omega}_{\gamma} D_{\gamma i}.
\end{equation}
Note that, from the decomposition of $\mathbf{D}$, Eq. \eqref{eqn:DDecomp}, $\bm{\rho}$ is parallel to the tangent line of the disclination. This is also the case for Eq. \eqref{eqn:DiscDensT} since $d \mathbf{R} / d s$ is directed along the tangent line. If $\mathbf{Q}$ is time dependent, then $\bm{\rho}$ and $\mathbf{D}$ are as well and $\mathbf{D}$ satisfies a continuity equation $(1/2)\partial_t D_{\gamma i} = \partial_{k} J_{\gamma i k}$ with
\begin{equation} \label{eqn:JDef}
    J_{\gamma i k} = \varepsilon_{i k \ell} \varepsilon_{\gamma \mu \nu} \partial_t Q_{\mu \alpha} \partial_{\ell} Q_{\nu \alpha}.
\end{equation}
Similar to Mazenko,\cite{mazenko97,mazenko99} we write $J_{\gamma i k} = \varepsilon_{i k \ell} g_{\gamma \ell}$ which will prove useful for analytic computations in the next section. 

To derive a continuity equation for $\bm{\rho}$, we first note that multiplying the continuity equation for $\mathbf{D}$ by a delta function gives $(1/2)D_{\gamma i} \partial_t \delta \left[\mathbf{Q}_{\perp}\right] = J_{\gamma i k} \partial_k \delta \left[\mathbf{Q}_{\perp} \right]$ which follows from properties of delta functions. With this result in hand, taking a time derivative of Eq. \eqref{eqn:DiscDensD} yields
\begin{equation} \label{eqn:ContE}
    \partial_t \rho_i = \partial_k \left( \delta\left[\mathbf{Q}_{\perp}\right] \hat{\Omega}_{\gamma} J_{\gamma i k} \right).
\end{equation}
Note that, as was the case in Section 2, even if $\bm{\hat{\Omega}}$ is time dependent, the term proportional to $\partial_t \bm{\hat{\Omega}}$ is zero since $\bm{\hat{\Omega}}$ is a unit vector so its time derivative is orthogonal to itself, thus $\partial_t \bm{\hat{\Omega}} \cdot \mathbf{D} = 0$. Eq. \eqref{eqn:ContE} is the standard result for topological continuity equations and the interpretation is that the disclination current is the topological density current restricted to the element of defect line. This result is important because we can write $J_{\gamma i k}$ in terms of the nematic order parameter. Additionally, connecting Eq. \eqref{eqn:ContE} with the time derivative of Eq. \eqref{eqn:DiscDensT} allows us to connect the velocity of the defect along the line, $\mathbf{v}(s)$, with the topological density current. \textcolor{black}{We find $\partial_t \rho_i = \partial_k\left(v_i \rho_k - v_k \rho_i\right)$ so that 
\begin{equation}
    \hat{\Omega}_{\tau} J_{\tau i k} = \frac{1}{2}\hat{\Omega}_{\gamma}\left(v_i D_{\gamma k} - v_k D_{\gamma i}\right).
\end{equation}}
This leads to $v_i = 2\hat{\Omega}_{\gamma} D_{\gamma k} \hat{\Omega}_{\tau} J_{\tau i k} / |\mathbf{D}|^2$ or, substituting Eq. \eqref{eqn:DDecomp},
\begin{equation} \label{eqn:Velocity}
    \mathbf{v}(s) = \left. 2\frac{\mathbf{\hat{T}} \times \left(\bm{\hat{\Omega}} \cdot \mathbf{g}\right)}{\omega}\right\rvert_{\mathbf{r} = \mathbf{R}(s)}
\end{equation}
with
\begin{equation} \label{eqn:gDef}
    g_{\gamma k} = \varepsilon_{\gamma \mu \nu} \partial_t Q_{\mu \alpha} \partial_k Q_{\nu \alpha}.
\end{equation}

Eq. \eqref{eqn:Velocity} is the second key result of this paper. Note that the velocity as written is explicitly orthogonal to the tangent vector of the disclination. Also of note is that $\mathbf{g}$ depends on the time evolution of $\mathbf{Q}$, but does not explicitly reference the source of the time evolution. That is, this expression is equally valid for nematics undergoing relaxational dynamics or active nematics with mass transport. To use this equation, one needs only to supply the appropriate time derivative of $\mathbf{Q}$. In the next section we use this result to predict the velocities analytically for a few examples of passive disclination lines and loops.

\section{Disclination velocity calculations for passive nematics}
Here we apply the formula for the velocity of a disclination, Eq. \eqref{eqn:Velocity}, to a few examples in passive nematics. To do this, we first make some simplifying assumptions. To compute $\mathbf{g}$, Eq. \eqref{eqn:gDef}, we must know $\mathbf{Q}$ and its spatial and time derivatives. We assume its time evolution is simply relaxational, $\partial_t \mathbf{Q} = -\Gamma \delta F / \delta \mathbf{Q}$, where $\Gamma$ is a rotational diffusion constant. We also assume that the free energy, $F$, has a functional derivative whose bulk term is analytic in $\mathbf{Q}$ at the core (such as the Landau-de Gennes free energy\cite{mottram14}), and has a one-constant elastic free energy. The former condition ensures that the only terms that survive in the calculation of $\mathbf{g}$ are those associated with the elastic energy, since it can be shown that $\varepsilon_{\gamma \mu \nu} \textcolor{black}{(}Q^n\textcolor{black}{)}_{\mu \alpha} \partial_k Q_{\nu \alpha} = 0$ at the core of a defect for any \textcolor{black}{power} $n$. Thus we can write $\partial_t \mathbf{Q} = \Gamma L \nabla^2 \mathbf{Q}$ where $L$ is the elastic constant.  

\begin{figure}[h]
\centering
  \includegraphics[height=6cm]{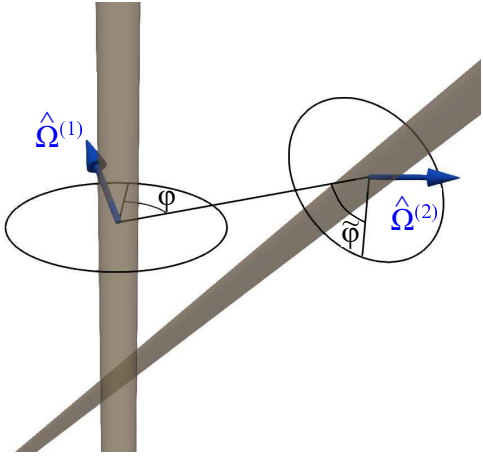}
  \caption{Interacting disclination lines at the closest point between the disclinations, represented by the line between them. $\varphi$ represents the azimuthal angle in the normal plane of the first disclination, while $\tilde{\varphi}$ represents the same for the second. $\bm{\hat{\Omega}}^{(1)}$ and $\bm{\hat{\Omega}}^{(2)}$ are assumed to be arbitrary for the calculation.}
  \label{fgr:InteractingDiscs}
\end{figure}

We begin with a calculation for interacting defect lines. Specifically, we calculate the velocity of the closest point between two disclinations. We assume the lines are straight and have constant rotation vectors which are otherwise arbitrary, and choose our axes so that the first disclination has tangent vector $\mathbf{\hat{T}}^{(1)} = \mathbf{\hat{z}}$, while the tangent vector for the second disclination is left arbitrary. This set up is sketched in Fig. \ref{fgr:InteractingDiscs} where the line segment connects the closest points of the disclinations. 

We first present an approximate, analytic calculation for the line velocity. For both lines, the director field near the disclination core is given by Eq. \eqref{eqn:nDisc} with a small perturbation occurring from the director field of the other. We define the two azimuthal angles to be zero along the line segment orthogonally connecting the lines. We denote the azimuth of the second disclination as $\tilde{\varphi}$ so that the perturbation of $\mathbf{\hat{n}}$ near the first disclination is expanded around $\tilde{\varphi} = 0$:
\begin{equation} \label{eqn:nPert}
    \mathbf{\hat{n}} \approx \left(\mathbf{\hat{n}_0} + \frac{1}{2}\tilde{\varphi} \mathbf{\hat{p}_0}\right) \cos \frac{1}{2}\varphi + \left(\mathbf{\hat{n}_1} + \frac{1}{2}\tilde{\varphi} \mathbf{\hat{p}_1} \right) \sin \frac{1}{2}\varphi
\end{equation}
where we define $\mathbf{\hat{p}_j} \equiv \bm{\hat{\Omega}}^{(2)} \times \mathbf{\hat{n}_j}$ and $\bm{\hat{\Omega}}^{(2)}$ is the rotation vector of the second disclination. The effect of the perturbation is to rotate the director slightly around $\bm{\hat{\Omega}}^{(2)}$ near the first disclination. Eq. \eqref{eqn:nPert} is then used with a linear core approximation\cite{long21} to yield an effective $\mathbf{Q}$ near the disclination,
\begin{multline} \label{eqn:QLineCore}
    \mathbf{Q} \approx S_0 \bigg[ \frac{1}{6} \mathbf{I} - \frac{1}{2} \boldsymbol{\hat{\Omega}}^{(1)} \otimes \boldsymbol{\hat{\Omega}}^{(1)} + \frac{1}{2} \frac{x}{a} \left( \mathbf{\tilde{n}_0} \otimes \mathbf{\tilde{n}_0} - \mathbf{\tilde{n}_1} \otimes \mathbf{\tilde{n}_1} \right)  \\ + \frac{1}{2} \frac{y}{a} \left( \mathbf{\tilde{n}_0} \otimes \mathbf{\tilde{n}_1} + \mathbf{\tilde{n}_1} \otimes \mathbf{\tilde{n}_0} \right) \bigg]
\end{multline}
where $a$ is the core radius and $\mathbf{\tilde{n}_j} \equiv \mathbf{\hat{n}_j} + (1/2)\tilde{\varphi} \mathbf{\hat{p}_j}$.

Equation \eqref{eqn:QLineCore} is particularly useful because the vectors $\mathbf{\hat{n}_j}$ and $\mathbf{\hat{p}_j}$ are constant. We compute $\mathbf{g}$, keeping in mind the fact that $\{\mathbf{\hat{n}_0},\mathbf{\hat{n}_1},\bm{\hat{\Omega}}^{(1)}\}$ form an orthogonal triad so that $\bm{\hat{\Omega}}^{(2)}$ can be written as a linear combination of the three. The result is
\begin{equation} \label{eqn:gLine}
    \left. \varepsilon_{\gamma \mu \nu} (\Gamma L \nabla^2 Q_{\mu \alpha}) \partial_k Q_{\nu \alpha} \right\rvert_{\mathbf{r} = \mathbf{0}} = -\frac{\Gamma L S_0^2}{a^2}\hat{\Omega}^{(2)}_{\gamma} \partial_k \tilde{\varphi} + \hat{n}_{0\gamma} A_k + \hat{n}_{1\gamma} B_k
\end{equation}
where $\mathbf{A}$ and $\mathbf{B}$ are vectors containing derivatives of $\tilde{\varphi}$ but are inconsequential to the result since $\mathbf{\hat{n}_0}$ and $\mathbf{\hat{n}_1}$ are orthogonal to $\bm{\hat{\Omega}}^{(1)}$. Using $\nabla \tilde{\varphi} = (\mathbf{\hat{T}}^{(2)} \times \mathbf{R}) / |\mathbf{R}|^2$ where $\mathbf{R}$ is the vector between the two disclinations, (i.e. $\mathbf{R} = \mathbf{r}^{(1)} - \mathbf{r}^{(2)}$) we find that the velocity of the closest point between disclinations is
\begin{equation} \label{eqn:LineVelocity}
    \mathbf{v} = 2 \Gamma L S_0^2 \left(\bm{\hat{\Omega}}^{(1)} \cdot \bm{\hat{\Omega}}^{(2)}\right) \left(\mathbf{\hat{T}}^{(1)} \cdot \mathbf{\hat{T}}^{(2)}\right)\frac{\mathbf{R}}{|\mathbf{R}|^2}.
\end{equation}

Equation \eqref{eqn:LineVelocity} indicates that if either the rotation vectors, or the tangent vectors of the two disclinations, are mutually orthogonal, the velocity vanishes. This qualitative result was also obtained by Long \textit{et al.}\cite{long21} in which the effective Peach-Koehler force was applied to a similar configuration. In fact, Eq. \eqref{eqn:LineVelocity} is identical to their result of the force between the two disclinations at their closest point up to coefficients. This suggests that Eq. \eqref{eqn:Velocity} for the velocity may be used to generalize the expression for the Peach-Koehler force to configurations with anisotropic elasticity or non-optimal orientation. \textcolor{black}{We note that here we have assumed the orientation to be optimal in approximating $\mathbf{Q}$ near the disclination core. By optimal, we mean that there is no relative difference between the extra degree of freedom of polarity of the defects (see Refs.\cite{vromans16,tang17}). However, Eq. \eqref{eqn:Velocity} holds regardless of this difference and the challenge of predicting the motion of defects in this case is in finding an accurate description of $\mathbf{Q}$ at the core.}

\begin{figure*}[h]
\centering
  \includegraphics[height=3.75in]{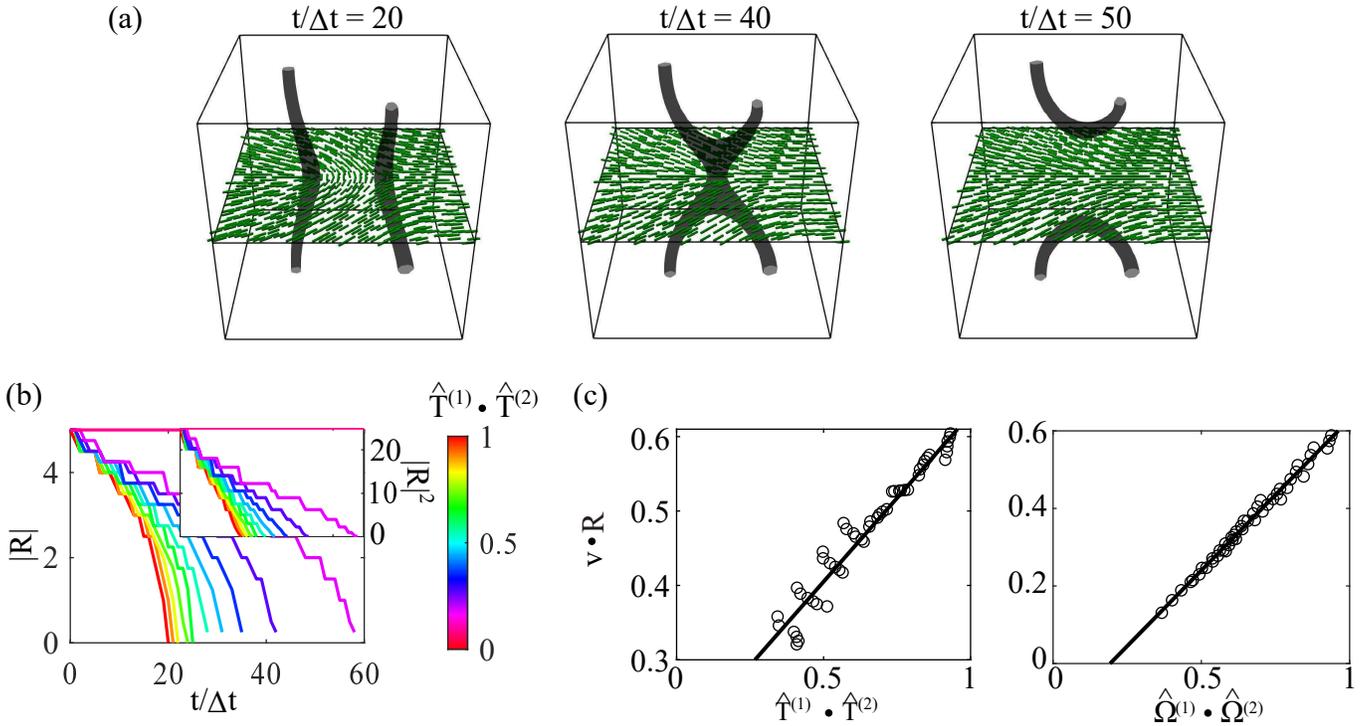}
  \caption{Analysis of recombining disclination lines. (a) Snapshots of the recombination of disclination lines with initial $\mathbf{\hat{T}}^{(1)}\cdot \mathbf{\hat{T}}^{(2)} = 0.3$ and $\bm{\hat{\Omega}}^{(1)} \cdot\bm{\hat{\Omega}}^{(2)} = -1$ at iteration numbers $t/\Delta t = 20,\,40,$ and $50$. Contours represent surfaces of $S = 0.3S_0$. Early in the computation, bends in the defects form near the closest points while late in the calculation, after the closest points annihilate, horseshoe shaped domains continue to recombine. (b) Distance of closest points, $|\mathbf{R}|$ versus iteration number for initial $\mathbf{\hat{T}}^{(1)}\cdot\mathbf{\hat{T}}^{(2)}$ varying from 0--1. The inset shows $|\mathbf{R}|^2$ versus iteration number, indicating a scaling $|\mathbf{R}| \sim t^{1/2}$ as predicted by Eq. \eqref{eqn:LineVelocity}. (c) \textcolor{black}{Instantaneous $\mathbf{v}\cdot\mathbf{R}$ versus instantaneous $\mathbf{\hat{T}}^{(1)}\cdot\mathbf{\hat{T}}^{(2)}$ with $\bm{\hat{\Omega}}^{(1)}\cdot\bm{\hat{\Omega}}^{(2)} = -1$ (left) and instantaneous $\bm{\hat{\Omega}}^{(1)}\cdot\bm{\hat{\Omega}}^{(2)}$ with $\mathbf{\hat{T}}^{(1)}\cdot\mathbf{\hat{T}}^{(2)} = 1$ (right) for calculations with various initial $\mathbf{\hat{T}}^{(1)}\cdot\mathbf{\hat{T}}^{(2)}$ and $\bm{\hat{\Omega}}^{(1)}\cdot\bm{\hat{\Omega}}^{(2)}$. Eq. \eqref{eqn:LineVelocity} predicts a linear scaling between these quantities which is confirmed by the numerical analysis. Note that the left plot only considers points where the line defects have no curvature since this is the valid regime for Eq. \eqref{eqn:LineVelocity}.}}
  \label{fgr:LineRecombo}
\end{figure*}

To check this calculation, we perform three-dimensional numerical calculations of recombination of disclination lines. We assume relaxational dynamics with a Maier-Saupe bulk potential\cite{ball10,schimming20,schimming20b} and a one-constant elastic free energy. The calculations are performed by using the finite element Matlab/C++ package FELICITY\cite{walker18} with a time dependent gradient flow algorithm\cite{schimming21} and an algebraic multi-grid linear equations solver.\cite{notay10,napov11,napov12,notay12} We cast the system in dimensionless units by defining the length scale $\xi = k_B T/L$ and time scale $\tau = 1/\Gamma k_B T$ and work in dimensionless position and time. We set the temperature so the liquid crystal is in the nematic phase ($S_0 = 0.6751$) and set the time-step $\Delta t = 0.1$. A standard tetrahedral mesh with $41\times41\times41$ vertices is used \textcolor{black}{with Neumann conditions at the boundary for all numerical calculations}. The closest point between disclinations is initialized to be $|\mathbf{R}| = 5$. \textcolor{black}{We perform computations varying the initial $\mathbf{\hat{T}}^{(1)}\cdot\mathbf{\hat{T}}^{(2)}$ from $0\text{--}1$ with $\bm{\hat{\Omega}}^{(1)}\cdot\bm{\hat{\Omega}}^{(2)} = -1$ as well as varying the initial $\bm{\hat{\Omega}}^{(1)}\cdot\bm{\hat{\Omega}}^{(2)}$ from $0 \text{--} 1$ with $\mathbf{\hat{T}}^{(1)}\cdot\mathbf{\hat{T}}^{(2)} = 1$. This allows us to independently analyze the effects of $\mathbf{\hat{T}}^{(1)}\cdot\mathbf{\hat{T}}^{(2)}$ and $\bm{\hat{\Omega}}^{(1)}\cdot\bm{\hat{\Omega}}^{(2)}$ on the velocity of the lines.}

Fig. \ref{fgr:LineRecombo}a shows several snapshots of the recombination process of two disclination lines. The contours represent surfaces of constant $S=0.3S_0$. Before the closest points annihilate, the disclination bends towards this point since the velocity is greatest between the two defects here. After the closest point annihilates, the remainder of the lines form two horseshoe shaped domains that continue to recombine until the texture disappears. Fig. \ref{fgr:LineRecombo}b shows $|\mathbf{R}|$ as a function of iteration number, $t / \Delta t$, for varying initial relative orientations of disclinations. The inset plots $|\mathbf{R}|^2$ vs iteration number which is linear for all events, indicating that $|\mathbf{R}| \sim t^{1/2}$, as expected from Eq. \eqref{eqn:LineVelocity}. Note that the fastest recombination is when the disclinations are parallel, while recombination never occurs when disclinations are perpendicular. \textcolor{black}{We find similar results in calculations with varying initial $\bm{\hat{\Omega}}^{(1)}\cdot\bm{\hat{\Omega}}^{(2)}$. In Fig. \ref{fgr:LineRecombo}c we show the instantaneous $\mathbf{v}\cdot\mathbf{R}$ as a function of instantaneous $\mathbf{\hat{T}}^{(1)}\cdot\mathbf{\hat{T}}^{(2)}$ and $\bm{\hat{\Omega}}^{(1)}\cdot\bm{\hat{\Omega}}^{(2)}$. As predicted by Eq. \eqref{eqn:LineVelocity}, we see a linear relationship between the two. We note that since Eq. \eqref{eqn:LineVelocity} only applies to straight defect lines and since the defect lines develop curvature relatively early into their evolution, the left plot of Fig. \ref{fgr:LineRecombo}c only shows points for the early evolution of the defects. As the defects develop curvature, the annihilation slows even though the tangent vectors are becoming more parallel due to an aligning torque. We also note that when numerically computing the velocity using Eq. \eqref{eqn:Velocity} at other points besides the closest points between defects we find velocities that qualitatively reflect the motion in the numerical calculations. A quantitative comparison would necessitate a finer mesh to more accurately compute the derivatives of the order parameter.}

We now calculate the velocity for a disclination loop. We will first focus on a loop with zero point charge, by which we mean that upon covering the loop with a measuring \textit{surface} there is not an overall covering of the order parameter space. Such loops have recently been the focus of three-dimensional active nematics since they are the fundamental excitations.\cite{duclos20,binysh20} For these loops the rotation vector is typically treated as being constant along the loop. In a passive nematic, there are two ``forces'' acting on a zero charge disclination loop. The first is the usual interaction between oppositely charged defects since at opposite ends of the loop $\bm{\hat{\Omega}} \cdot \mathbf{\hat{T}}$ changes sign (or handedness for the case of the twist deformation when $\bm{\hat{\Omega}} \cdot \mathbf{\hat{T}} = 0$). The second force seeks to make the defect as small as possible due to the self energy of the defect core. Hence, for a zero charge disclination loop, both forces act to annihilate the defect. 

Here, we will only compute the velocity due to the self energy as a careful consideration of the interaction induced by other parts of the defect is beyond the scope of this work. To compute this, we take $\mathbf{\hat{n}}$ as in Eq. \eqref{eqn:nDisc} and work in cylindrical coordinates, assuming the loop lies in the $xy$-plane. Taking a linear core approximation, $\mathbf{Q}$ near the core is given by
\begin{multline} \label{eqn:QLoopCore}
   \mathbf{Q} = S_0 \bigg[ \frac{1}{6} \mathbf{I} - \frac{1}{2} \boldsymbol{\hat{\Omega}} \otimes \boldsymbol{\hat{\Omega}} + \frac{1}{2} \frac{\rho - R}{a} \left( \mathbf{\hat{n}_0} \otimes \mathbf{\hat{n}_0} - \mathbf{\hat{n}_1} \otimes \mathbf{\hat{n}_1} \right)  \\ + \frac{1}{2} \frac{z}{a} \left( \mathbf{\hat{n}_0} \otimes \mathbf{\hat{n}_1} + \mathbf{\hat{n}_1} \otimes \mathbf{\hat{n}_0} \right) \bigg].
\end{multline}
where $R$ is the disclination radius. We assume the triad $\{\mathbf{\hat{n}_0},\,\mathbf{\hat{n}_1},\,\bm{\hat{\Omega}}\}$ is constant throughout the loop and so we find
\begin{equation} \label{eqn:loopVelocity}
    \mathbf{v} = -\frac{\Gamma L S_0^2}{R}\bm{\hat{\rho}},
\end{equation}
indicating that the loop is shrinking at a rate inversely proportional to $R$. Note that the interaction with the rest of loop should scale similarly since the interaction is Coulomb-like.\cite{deGennes75} This means the effect of the interaction only changes the coefficient in front of Eq. \eqref{eqn:loopVelocity} and the qualitative results hold. Thus we expect the time dependence of the radius to be $R(t) \sim t^{1/2}$, the same time dependence as the distance between two annihilating defects in two dimensions.\cite{zumer02} Because this loop has zero point charge, it can self-annihilate and the configuration will eventually become uniform.

\begin{figure}[h]
\centering
  \includegraphics[height=4.25in]{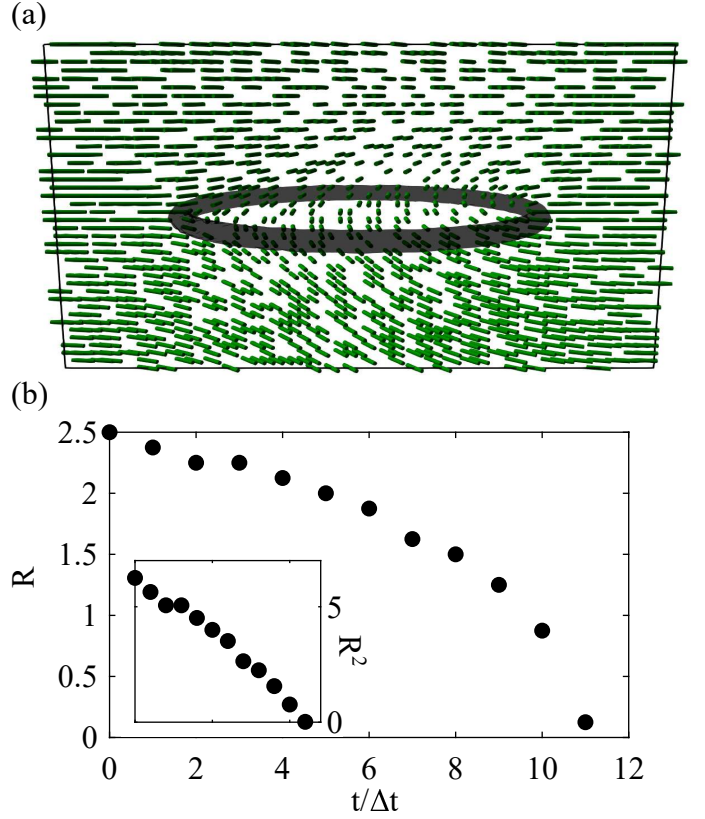}
  \caption{Self-annihilating pure twist loop. (a) Snapshot at $t/\Delta t = 5$ of a pure twist disclination loop with $\bm{\hat{\Omega}}=\mathbf{\hat{z}}$ along the loop. The contour represents a surface of constant $S=0.3S_0$. (b) Loop radius $R$ plotted against iteration number. The inset shows $R^2$ versus iteration number, demonstrating the scaling $R \sim t^{1/2}$.}
  \label{fgr:PureTwistLoop}
\end{figure}

Fig. \ref{fgr:PureTwistLoop} shows results of a numerical calculation for a zero charge loop. Specifically, we simulate a pure-twist loop where $\bm{\hat{\Omega}}\cdot\mathbf{\hat{T}} = 0$ everywhere along the loop. The details of the numerics are similar to those outlined above except here we use a tetrahedral mesh with $81\times81\times81$ vertices. Fig. \ref{fgr:PureTwistLoop}b shows the time dependence of the radius, $R(t)$, while the inset shows $R^2(t)$. The time dependence demonstrates that the radius scales as $R(t) \sim t^{1/2}$ which is expected from Eq. \eqref{eqn:loopVelocity}. We note that, for a one-constant elastic energy, the value $\bm{\hat{\Omega}}$ does not change the dynamics of the loop. We also note that the loop shrinks much faster than the pair of disclinations studied in Fig. \ref{fgr:LineRecombo}. The initial diameter of the loop is the same as the initial distance between disclination lines, yet upon comparing the recombination time we see the disclination loop annihilates at time-step $12$ while the parallel disclination lines annihilate at time-step $20$. \textcolor{black}{This behavior is inconsistent with the calculated coefficients in Eqs. \eqref{eqn:LineVelocity} and \eqref{eqn:loopVelocity}; however,} this is likely due to the fact that there are two forces acting on the disclination loop as opposed to one on the disclination lines \textcolor{black}{and the computation above only reflects one such force as discussed}. We further note that both the $t^{1/2}$ scaling for lines and loops, and the qualitative features, such as the bends in the disclination lines and the remaining horseshoe structures have been recently observed experimentally by Zushi and Takeuchi.\cite{zushi21}

We conclude this section with a qualitative prediction for the velocity of a \textit{nonzero} point charge loop defect. Loops of this nature are typically associated with colloidal particles with homeotropic boundary conditions.\cite{gu00,stark01,alama16,tovkach17} Here, we assume there is no particle and there is only a loop with $\bm{\hat{\Omega}}\cdot\mathbf{\hat{T}} = 1$, which, in the far field, appears as a radial hedgehog point defect. As above, we will only compute the contribution from the self energy. Here, one must be careful since the triad $\{\mathbf{\hat{n}_0},\,\mathbf{\hat{n}_1},\,\bm{\hat{\Omega}}\}$ changes along the loop. To do the calculation we assume $\mathbf{\hat{n}_0} = \bm{\hat{\rho}}$ and $\bm{\hat{\Omega}} = \mathbf{\hat{T}} = -\bm{\hat{\phi}}$ so that $\partial \mathbf{\hat{n}_0} / \partial \phi = -\bm{\hat{\Omega}}$ and $\partial \bm{\hat{\Omega}} / \partial \phi =  \mathbf{\hat{n}_0}$. By using these relations we arrive at
\begin{equation}
    \mathbf{v} = \Gamma L S_0^2 \left( -\frac{1}{R} + \frac{a}{R^2} \right) \bm{\hat{\rho}}.
\end{equation}
Because there is an overall point charge one might assume that the defect loop would shrink and become a typical hedgehog defect. However, this result shows that there is a stable size for the loop instead. If the loop is much larger than this size, the nonzero charge loop behaves similarly to the zero charge loop. However, if the loop is smaller than this size, it grows rapidly. Since we are only considering the self energy of the defect, the physical interpretation is that a small loop has a smaller energy than a point defect core. This stability of a loop defect over a point defect has been seen in previous numerical and analytic studies of hedgehog defect cores.\cite{stark01,tovkach17} This is similar to the case in two-dimensions where a single $\pm 1$ defect splits into two $\pm 1/2$ defects to lower the energy. However, in that case, the defects repel each other and are only stabilized by other defects or boundary conditions.\cite{kim13,vromans16} Of course, if the interaction with other parts of the loop are taken into account, we expect that the coefficient on the $1/R$ term would change. Thus, one would need to perform a more detailed calculation for a quantitative prediction of the size of the stable loop.

\section{Conclusion}
In this work we have extended recent efforts to fundamentally understand the nature of defect lines and loops in nematics. We have introduced a disclination density tensor, $\mathbf{D}$, that can be computed from first derivatives of the tensor order parameter and is nonzero at defect locations. This tensor decomposes as a dyadic combination of unit vectors that geometricly define the disclination. We have derived a continuity equation for the topological charge, and explicitly written a velocity for the defect line. Further, we have demonstrated with several examples the practicality of the velocity equation, Eq. \eqref{eqn:Velocity}, by analytically deriving results for different disclination configurations.

There is still more work to be done in understanding disclination dynamics. \textcolor{black}{As demonstrated by Eq. \eqref{eqn:Velocity}, the velocity of a line disclination depends on its instantaneous rotation vector, $\bm{\hat{\Omega}}$. The issue of understanding how the rotation vector evolves in time remains a challenge. Recent theory and experiment\cite{houston21,zushi21} have begun to explore this issue.} Moreover, it will be interesting to see how Eq. \eqref{eqn:Velocity} can be applied to systems with a more complex time dependence, either for active systems, systems with anisotropic elasticity or both. 


\section*{Conflicts of interest}
There are no conflicts to declare.

\section*{Acknowledgements}
This research has been supported by the National Science Foundation under Grant No. DMR-1838977, and by the Minnesota Supercomputing Institute.





\bibliography{LC} 

\providecommand*{\mcitethebibliography}{\thebibliography}
\csname @ifundefined\endcsname{endmcitethebibliography}
{\let\endmcitethebibliography\endthebibliography}{}
\begin{mcitethebibliography}{48}
\providecommand*{\natexlab}[1]{#1}
\providecommand*{\mciteSetBstSublistMode}[1]{}
\providecommand*{\mciteSetBstMaxWidthForm}[2]{}
\providecommand*{\mciteBstWouldAddEndPuncttrue}
  {\def\EndOfBibitem{\unskip.}}
\providecommand*{\mciteBstWouldAddEndPunctfalse}
  {\let\EndOfBibitem\relax}
\providecommand*{\mciteSetBstMidEndSepPunct}[3]{}
\providecommand*{\mciteSetBstSublistLabelBeginEnd}[3]{}
\providecommand*{\EndOfBibitem}{}
\mciteSetBstSublistMode{f}
\mciteSetBstMaxWidthForm{subitem}
{(\emph{\alph{mcitesubitemcount}})}
\mciteSetBstSublistLabelBeginEnd{\mcitemaxwidthsubitemform\space}
{\relax}{\relax}

\bibitem[Chaikin and Lubensky(1995)]{chaikin95}
P.~M. Chaikin and T.~C. Lubensky, \emph{Principles of Condensed Matter
  Physics}, Cambridge University Press, 1995\relax
\mciteBstWouldAddEndPuncttrue
\mciteSetBstMidEndSepPunct{\mcitedefaultmidpunct}
{\mcitedefaultendpunct}{\mcitedefaultseppunct}\relax
\EndOfBibitem
\bibitem[Kibble(1997)]{kibble97}
T.~W.~B. Kibble, \emph{Aust. J. Phys.}, 1997, \textbf{50}, 697\relax
\mciteBstWouldAddEndPuncttrue
\mciteSetBstMidEndSepPunct{\mcitedefaultmidpunct}
{\mcitedefaultendpunct}{\mcitedefaultseppunct}\relax
\EndOfBibitem
\bibitem[Pismen(1999)]{pismen99}
L.~M. Pismen, \emph{Vortices in Nonlinear Fields}, Oxford University Press,
  1999\relax
\mciteBstWouldAddEndPuncttrue
\mciteSetBstMidEndSepPunct{\mcitedefaultmidpunct}
{\mcitedefaultendpunct}{\mcitedefaultseppunct}\relax
\EndOfBibitem
\bibitem[Friedel and De~Gennes(1969)]{friedel69}
J.~Friedel and P.~De~Gennes, \emph{CR Acad. Sc. Paris B}, 1969, \textbf{268},
  257--259\relax
\mciteBstWouldAddEndPuncttrue
\mciteSetBstMidEndSepPunct{\mcitedefaultmidpunct}
{\mcitedefaultendpunct}{\mcitedefaultseppunct}\relax
\EndOfBibitem
\bibitem[de~Gennes(1975)]{deGennes75}
P.~G. de~Gennes, \emph{The Physics of Liquid Crystals}, Oxford University
  Press, 1975\relax
\mciteBstWouldAddEndPuncttrue
\mciteSetBstMidEndSepPunct{\mcitedefaultmidpunct}
{\mcitedefaultendpunct}{\mcitedefaultseppunct}\relax
\EndOfBibitem
\bibitem[Kim \emph{et~al.}(2013)Kim, Shiyanovskii, and Lavrentovich]{kim13}
Y.~K. Kim, S.~V. Shiyanovskii and O.~D. Lavrentovich, \emph{J. Phys.: Condens.
  Matter}, 2013, \textbf{25}, 404202\relax
\mciteBstWouldAddEndPuncttrue
\mciteSetBstMidEndSepPunct{\mcitedefaultmidpunct}
{\mcitedefaultendpunct}{\mcitedefaultseppunct}\relax
\EndOfBibitem
\bibitem[Alexander \emph{et~al.}(2012)Alexander, Chen, Matsumoto, and
  Kamien]{alexander12}
G.~P. Alexander, B.~G.-g. Chen, E.~Matsumoto and R.~D. Kamien, \emph{Rev. Mod.
  Phys.}, 2012, \textbf{84}, 497\relax
\mciteBstWouldAddEndPuncttrue
\mciteSetBstMidEndSepPunct{\mcitedefaultmidpunct}
{\mcitedefaultendpunct}{\mcitedefaultseppunct}\relax
\EndOfBibitem
\bibitem[Gu and Abbott(2000)]{gu00}
Y.~Gu and N.~L. Abbott, \emph{Phys. Rev. Lett.}, 2000, \textbf{85}, 4719\relax
\mciteBstWouldAddEndPuncttrue
\mciteSetBstMidEndSepPunct{\mcitedefaultmidpunct}
{\mcitedefaultendpunct}{\mcitedefaultseppunct}\relax
\EndOfBibitem
\bibitem[Stark(2001)]{stark01}
H.~Stark, \emph{Physics Reports}, 2001, \textbf{351}, 387--474\relax
\mciteBstWouldAddEndPuncttrue
\mciteSetBstMidEndSepPunct{\mcitedefaultmidpunct}
{\mcitedefaultendpunct}{\mcitedefaultseppunct}\relax
\EndOfBibitem
\bibitem[Alama \emph{et~al.}(2016)Alama, Bronsard, and Lamy]{alama16}
S.~Alama, L.~Bronsard and X.~Lamy, \emph{Phys. Rev. E}, 2016, \textbf{93},
  012705\relax
\mciteBstWouldAddEndPuncttrue
\mciteSetBstMidEndSepPunct{\mcitedefaultmidpunct}
{\mcitedefaultendpunct}{\mcitedefaultseppunct}\relax
\EndOfBibitem
\bibitem[Mostajeran(2015)]{most15}
C.~Mostajeran, \emph{Phys. Rev. E}, 2015, \textbf{91}, 062405\relax
\mciteBstWouldAddEndPuncttrue
\mciteSetBstMidEndSepPunct{\mcitedefaultmidpunct}
{\mcitedefaultendpunct}{\mcitedefaultseppunct}\relax
\EndOfBibitem
\bibitem[Babakhanova \emph{et~al.}(2018)Babakhanova, Turiv, Guo, Hendrikx, Wei,
  Schenning, Broer, and Lavrentovich]{baba18}
G.~Babakhanova, T.~Turiv, Y.~Guo, M.~Hendrikx, Q.-H. Wei, A.~P. Schenning,
  D.~J. Broer and O.~D. Lavrentovich, \emph{Nat. Commun.}, 2018, \textbf{9},
  456\relax
\mciteBstWouldAddEndPuncttrue
\mciteSetBstMidEndSepPunct{\mcitedefaultmidpunct}
{\mcitedefaultendpunct}{\mcitedefaultseppunct}\relax
\EndOfBibitem
\bibitem[Marchetti \emph{et~al.}(2013)Marchetti, Joanny, Ramaswamy, Liverpool,
  Prost, Rao, and Simha]{marchetti13}
M.~Marchetti, J.~Joanny, S.~Ramaswamy, T.~Liverpool, J.~Prost, M.~Rao and R.~A.
  Simha, \emph{Rev. Mod. Phys.}, 2013, \textbf{85}, 1143\relax
\mciteBstWouldAddEndPuncttrue
\mciteSetBstMidEndSepPunct{\mcitedefaultmidpunct}
{\mcitedefaultendpunct}{\mcitedefaultseppunct}\relax
\EndOfBibitem
\bibitem[Ramaswamy(2017)]{ramaswamy17}
S.~Ramaswamy, \emph{J. Stat. Mech.}, 2017, \textbf{2017}, 054002\relax
\mciteBstWouldAddEndPuncttrue
\mciteSetBstMidEndSepPunct{\mcitedefaultmidpunct}
{\mcitedefaultendpunct}{\mcitedefaultseppunct}\relax
\EndOfBibitem
\bibitem[Aranson(2019)]{aranson19}
I.~S. Aranson, \emph{Phys.-Usp.}, 2019, \textbf{62}, 892\relax
\mciteBstWouldAddEndPuncttrue
\mciteSetBstMidEndSepPunct{\mcitedefaultmidpunct}
{\mcitedefaultendpunct}{\mcitedefaultseppunct}\relax
\EndOfBibitem
\bibitem[Kumar \emph{et~al.}(2018)Kumar, Zhang, De~Pablo, and Gardel]{kumar18}
N.~Kumar, R.~Zhang, J.~J. De~Pablo and M.~L. Gardel, \emph{Sci. Adv.}, 2018,
  \textbf{4}, eaat7779\relax
\mciteBstWouldAddEndPuncttrue
\mciteSetBstMidEndSepPunct{\mcitedefaultmidpunct}
{\mcitedefaultendpunct}{\mcitedefaultseppunct}\relax
\EndOfBibitem
\bibitem[Genkin \emph{et~al.}(2017)Genkin, Sokolov, Lavrentovich, and
  Aranson]{genkin17}
M.~M. Genkin, A.~Sokolov, O.~D. Lavrentovich and I.~S. Aranson, \emph{Phys.
  Rev. X}, 2017, \textbf{7}, 011029\relax
\mciteBstWouldAddEndPuncttrue
\mciteSetBstMidEndSepPunct{\mcitedefaultmidpunct}
{\mcitedefaultendpunct}{\mcitedefaultseppunct}\relax
\EndOfBibitem
\bibitem[Nishiguchi \emph{et~al.}(2017)Nishiguchi, Nagai, Chat{\'e}, and
  Sano]{nishiguchi17}
D.~Nishiguchi, K.~H. Nagai, H.~Chat{\'e} and M.~Sano, \emph{Phys. Rev. E},
  2017, \textbf{95}, 020601(R)\relax
\mciteBstWouldAddEndPuncttrue
\mciteSetBstMidEndSepPunct{\mcitedefaultmidpunct}
{\mcitedefaultendpunct}{\mcitedefaultseppunct}\relax
\EndOfBibitem
\bibitem[Copenhagen \emph{et~al.}(2021)Copenhagen, Alert, Wingreen, and
  Shaevitz]{copenhagen21}
K.~Copenhagen, R.~Alert, N.~S. Wingreen and J.~W. Shaevitz, \emph{Nat. Phys.},
  2021, \textbf{17}, 211--215\relax
\mciteBstWouldAddEndPuncttrue
\mciteSetBstMidEndSepPunct{\mcitedefaultmidpunct}
{\mcitedefaultendpunct}{\mcitedefaultseppunct}\relax
\EndOfBibitem
\bibitem[Saw \emph{et~al.}(2017)Saw, Doostmohammadi, Nier, Kocgozlu, Thampi,
  Toyama, Marcq, Lim, Yeomans, and Ladoux]{saw17}
T.~B. Saw, A.~Doostmohammadi, V.~Nier, L.~Kocgozlu, S.~Thampi, Y.~Toyama,
  P.~Marcq, C.~T. Lim, J.~M. Yeomans and B.~Ladoux, \emph{Nature}, 2017,
  \textbf{544}, 212--216\relax
\mciteBstWouldAddEndPuncttrue
\mciteSetBstMidEndSepPunct{\mcitedefaultmidpunct}
{\mcitedefaultendpunct}{\mcitedefaultseppunct}\relax
\EndOfBibitem
\bibitem[Duclos \emph{et~al.}(2020)Duclos, Adkins, Banerjee, Peterson,
  Varghese, Kolvin, Baskaran, Pelcovits, Powers, Baskaran, Toschi, Hagan,
  Streichan, Vitelli, Beller, and Dogic]{duclos20}
G.~Duclos, R.~Adkins, D.~Banerjee, M.~S.~E. Peterson, M.~Varghese, I.~Kolvin,
  A.~Baskaran, R.~A. Pelcovits, T.~R. Powers, A.~Baskaran, F.~Toschi, M.~F.
  Hagan, S.~J. Streichan, V.~Vitelli, D.~A. Beller and Z.~Dogic,
  \emph{Science}, 2020, \textbf{367}, 1120--1124\relax
\mciteBstWouldAddEndPuncttrue
\mciteSetBstMidEndSepPunct{\mcitedefaultmidpunct}
{\mcitedefaultendpunct}{\mcitedefaultseppunct}\relax
\EndOfBibitem
\bibitem[Binysh \emph{et~al.}(2020)Binysh, Kos, \u{C}opar, Ravnik, and
  Alexander]{binysh20}
J.~Binysh, u.~Kos, S.~\u{C}opar, M.~Ravnik and G.~P. Alexander, \emph{Phys.
  Rev. Lett.}, 2020, \textbf{124}, 088001\relax
\mciteBstWouldAddEndPuncttrue
\mciteSetBstMidEndSepPunct{\mcitedefaultmidpunct}
{\mcitedefaultendpunct}{\mcitedefaultseppunct}\relax
\EndOfBibitem
\bibitem[Houston and Alexander(2021)]{houston21}
A.~J. Houston and P.~Alexander, Gareth, \emph{Defect Loops in Three-Dimensional
  Active Nematics as Active Multipoles}, e-print
  arXiv:2106.15424[cond-mat.soft], 2021\relax
\mciteBstWouldAddEndPuncttrue
\mciteSetBstMidEndSepPunct{\mcitedefaultmidpunct}
{\mcitedefaultendpunct}{\mcitedefaultseppunct}\relax
\EndOfBibitem
\bibitem[Long \emph{et~al.}(2021)Long, Tang, Selinger, and Selinger]{long21}
C.~Long, X.~Tang, R.~L. Selinger and J.~V. Selinger, \emph{Soft Matter}, 2021,
  \textbf{17}, 2265\relax
\mciteBstWouldAddEndPuncttrue
\mciteSetBstMidEndSepPunct{\mcitedefaultmidpunct}
{\mcitedefaultendpunct}{\mcitedefaultseppunct}\relax
\EndOfBibitem
\bibitem[Shankar and Marchetti(2019)]{shankar19}
S.~Shankar and M.~C. Marchetti, \emph{Phys. Rev. X}, 2019, \textbf{9},
  041047\relax
\mciteBstWouldAddEndPuncttrue
\mciteSetBstMidEndSepPunct{\mcitedefaultmidpunct}
{\mcitedefaultendpunct}{\mcitedefaultseppunct}\relax
\EndOfBibitem
\bibitem[Angheluta \emph{et~al.}(2021)Angheluta, Chen, Marchetti, and
  Bowick]{angheluta21}
L.~Angheluta, Z.~Chen, M.~C. Marchetti and M.~J. Bowick, \emph{New J. Phys.},
  2021, \textbf{23}, 033009\relax
\mciteBstWouldAddEndPuncttrue
\mciteSetBstMidEndSepPunct{\mcitedefaultmidpunct}
{\mcitedefaultendpunct}{\mcitedefaultseppunct}\relax
\EndOfBibitem
\bibitem[Liu and Mazenko(1992)]{liu92}
F.~Liu and G.~F. Mazenko, \emph{Phys. Rev. B}, 1992, \textbf{46}, 5963\relax
\mciteBstWouldAddEndPuncttrue
\mciteSetBstMidEndSepPunct{\mcitedefaultmidpunct}
{\mcitedefaultendpunct}{\mcitedefaultseppunct}\relax
\EndOfBibitem
\bibitem[Mazenko and Wickham(1997)]{mazenko97}
G.~F. Mazenko and R.~A. Wickham, \emph{Phys. Rev. E}, 1997, \textbf{57},
  2539\relax
\mciteBstWouldAddEndPuncttrue
\mciteSetBstMidEndSepPunct{\mcitedefaultmidpunct}
{\mcitedefaultendpunct}{\mcitedefaultseppunct}\relax
\EndOfBibitem
\bibitem[Mottram and Newton(2014)]{mottram14}
N.~J. Mottram and C.~J. Newton, \emph{Introduction to {Q}-tensor theory},
  e-print arXiv:1409.3542v2 [cond-mat.soft], 2014\relax
\mciteBstWouldAddEndPuncttrue
\mciteSetBstMidEndSepPunct{\mcitedefaultmidpunct}
{\mcitedefaultendpunct}{\mcitedefaultseppunct}\relax
\EndOfBibitem
\bibitem[Halperin(1981)]{halperin81}
B.~I. Halperin, \emph{Physics of Defects}, North-Holland Pub. Co., 1981\relax
\mciteBstWouldAddEndPuncttrue
\mciteSetBstMidEndSepPunct{\mcitedefaultmidpunct}
{\mcitedefaultendpunct}{\mcitedefaultseppunct}\relax
\EndOfBibitem
\bibitem[Schopohl and Sluckin(1987)]{schopohl87}
N.~Schopohl and T.~Sluckin, \emph{Phys. Rev. Lett.}, 1987, \textbf{59},
  22\relax
\mciteBstWouldAddEndPuncttrue
\mciteSetBstMidEndSepPunct{\mcitedefaultmidpunct}
{\mcitedefaultendpunct}{\mcitedefaultseppunct}\relax
\EndOfBibitem
\bibitem[Blow \emph{et~al.}(2014)Blow, Thampi, and Yeomans]{blow14}
M.~L. Blow, S.~P. Thampi and J.~M. Yeomans, \emph{Phys. Rev. Lett.}, 2014,
  \textbf{113}, 248303\relax
\mciteBstWouldAddEndPuncttrue
\mciteSetBstMidEndSepPunct{\mcitedefaultmidpunct}
{\mcitedefaultendpunct}{\mcitedefaultseppunct}\relax
\EndOfBibitem
\bibitem[Dell'Arciprete \emph{et~al.}(2018)Dell'Arciprete, Blow, Brown,
  Farrell, Lintuvuori, McVey, Marenduzzo, and Poon]{dell18}
D.~Dell'Arciprete, M.~Blow, A.~Brown, F.~Farrell, J.~S. Lintuvuori, A.~F.
  McVey, D.~Marenduzzo and W.~Poon, \emph{Nat. Commun.}, 2018, \textbf{9},
  4190\relax
\mciteBstWouldAddEndPuncttrue
\mciteSetBstMidEndSepPunct{\mcitedefaultmidpunct}
{\mcitedefaultendpunct}{\mcitedefaultseppunct}\relax
\EndOfBibitem
\bibitem[Mazenko(1999)]{mazenko99}
G.~F. Mazenko, \emph{Phys. Rev. E}, 1999, \textbf{59}, 1574\relax
\mciteBstWouldAddEndPuncttrue
\mciteSetBstMidEndSepPunct{\mcitedefaultmidpunct}
{\mcitedefaultendpunct}{\mcitedefaultseppunct}\relax
\EndOfBibitem
\bibitem[Vromans and Giomi(2016)]{vromans16}
A.~J. Vromans and L.~Giomi, \emph{Soft Matter}, 2016, \textbf{12}, 6490\relax
\mciteBstWouldAddEndPuncttrue
\mciteSetBstMidEndSepPunct{\mcitedefaultmidpunct}
{\mcitedefaultendpunct}{\mcitedefaultseppunct}\relax
\EndOfBibitem
\bibitem[Tang and Selinger(2017)]{tang17}
X.~Tang and J.~V. Selinger, \emph{Soft Matter}, 2017, \textbf{13}, 5481\relax
\mciteBstWouldAddEndPuncttrue
\mciteSetBstMidEndSepPunct{\mcitedefaultmidpunct}
{\mcitedefaultendpunct}{\mcitedefaultseppunct}\relax
\EndOfBibitem
\bibitem[Ball and Majumdar(2010)]{ball10}
J.~M. Ball and A.~Majumdar, \emph{Mol. liq. Cryst.}, 2010, \textbf{525},
  1\relax
\mciteBstWouldAddEndPuncttrue
\mciteSetBstMidEndSepPunct{\mcitedefaultmidpunct}
{\mcitedefaultendpunct}{\mcitedefaultseppunct}\relax
\EndOfBibitem
\bibitem[Schimming and Vi\~{n}als(2020)]{schimming20}
C.~D. Schimming and J.~Vi\~{n}als, \emph{Phys. Rev. E.}, 2020, \textbf{101},
  032702\relax
\mciteBstWouldAddEndPuncttrue
\mciteSetBstMidEndSepPunct{\mcitedefaultmidpunct}
{\mcitedefaultendpunct}{\mcitedefaultseppunct}\relax
\EndOfBibitem
\bibitem[Schimming and Vi{\~n}als(2020)]{schimming20b}
C.~D. Schimming and J.~Vi{\~n}als, \emph{Phys. Rev. E}, 2020, \textbf{102},
  010701\relax
\mciteBstWouldAddEndPuncttrue
\mciteSetBstMidEndSepPunct{\mcitedefaultmidpunct}
{\mcitedefaultendpunct}{\mcitedefaultseppunct}\relax
\EndOfBibitem
\bibitem[Walker(2018)]{walker18}
S.~W. Walker, \emph{SIAM J. Sci. Comput.}, 2018, \textbf{40}, C234--C257\relax
\mciteBstWouldAddEndPuncttrue
\mciteSetBstMidEndSepPunct{\mcitedefaultmidpunct}
{\mcitedefaultendpunct}{\mcitedefaultseppunct}\relax
\EndOfBibitem
\bibitem[Schimming \emph{et~al.}(2021)Schimming, Vi\~{n}als, and
  Walker]{schimming21}
C.~D. Schimming, J.~Vi\~{n}als and S.~W. Walker, \emph{J. Comp. Phys.}, 2021,
  \textbf{441}, 110441\relax
\mciteBstWouldAddEndPuncttrue
\mciteSetBstMidEndSepPunct{\mcitedefaultmidpunct}
{\mcitedefaultendpunct}{\mcitedefaultseppunct}\relax
\EndOfBibitem
\bibitem[Notay(2010)]{notay10}
Y.~Notay, \emph{Electron. Trans. Numer. Anal.}, 2010, \textbf{37},
  123--146\relax
\mciteBstWouldAddEndPuncttrue
\mciteSetBstMidEndSepPunct{\mcitedefaultmidpunct}
{\mcitedefaultendpunct}{\mcitedefaultseppunct}\relax
\EndOfBibitem
\bibitem[Napov and Notay(2011)]{napov11}
A.~Napov and Y.~Notay, \emph{Numer. Linear Algebra Appl.}, 2011, \textbf{18},
  539--564\relax
\mciteBstWouldAddEndPuncttrue
\mciteSetBstMidEndSepPunct{\mcitedefaultmidpunct}
{\mcitedefaultendpunct}{\mcitedefaultseppunct}\relax
\EndOfBibitem
\bibitem[Napov and Notay(2012)]{napov12}
A.~Napov and Y.~Notay, \emph{SIAM J. Sci. Comput.}, 2012, \textbf{34},
  A1079--A1109\relax
\mciteBstWouldAddEndPuncttrue
\mciteSetBstMidEndSepPunct{\mcitedefaultmidpunct}
{\mcitedefaultendpunct}{\mcitedefaultseppunct}\relax
\EndOfBibitem
\bibitem[Notay(2012)]{notay12}
Y.~Notay, \emph{SIAM J. Sci. Comput.}, 2012, \textbf{34}, A2288--A2316\relax
\mciteBstWouldAddEndPuncttrue
\mciteSetBstMidEndSepPunct{\mcitedefaultmidpunct}
{\mcitedefaultendpunct}{\mcitedefaultseppunct}\relax
\EndOfBibitem
\bibitem[\u{Z}umer and Sven\u{s}ek(2002)]{zumer02}
S.~\u{Z}umer and D.~Sven\u{s}ek, \emph{Phys. Rev. E}, 2002, \textbf{66},
  021712\relax
\mciteBstWouldAddEndPuncttrue
\mciteSetBstMidEndSepPunct{\mcitedefaultmidpunct}
{\mcitedefaultendpunct}{\mcitedefaultseppunct}\relax
\EndOfBibitem
\bibitem[Zushi and Takeuchi(2021)]{zushi21}
Y.~Zushi and K.~A. Takeuchi, \emph{Scaling and Spontaneous Symmetry Restoring
  in Reconnecting Nematic Disclinations}, e-print arXiv:2110.00442
  [cond-mat.soft], 2021\relax
\mciteBstWouldAddEndPuncttrue
\mciteSetBstMidEndSepPunct{\mcitedefaultmidpunct}
{\mcitedefaultendpunct}{\mcitedefaultseppunct}\relax
\EndOfBibitem
\bibitem[Tovkach \emph{et~al.}(2017)Tovkach, Conklin, Calderer, Golovaty,
  Lavrentovich, {Vi\~nals}, and Walkington]{tovkach17}
O.~M. Tovkach, C.~Conklin, M.~C. Calderer, D.~Golovaty, O.~D. Lavrentovich,
  J.~{Vi\~nals} and N.~J. Walkington, \emph{Physical Review Fluids}, 2017,
  \textbf{2}, 053302\relax
\mciteBstWouldAddEndPuncttrue
\mciteSetBstMidEndSepPunct{\mcitedefaultmidpunct}
{\mcitedefaultendpunct}{\mcitedefaultseppunct}\relax
\EndOfBibitem
\end{mcitethebibliography}
\bibliographystyle{rsc} 

\end{document}